 \documentclass[sigconf]{acmart}
\AtBeginDocument{%
  \providecommand\BibTeX{{%
    \normalfont B\kern-0.5em{\scshape i\kern-0.25em b}\kern-0.8em\TeX}}}

\setcopyright{acmlicensed}
\copyrightyear{2025}
\acmYear{2025}
\setcopyright{rightsretained}
\acmConference[CHI EA '25]{Extended Abstracts of the CHI Conference on Human Factors in Computing Systems}{April 26-May 1, 2025}{Yokohama, Japan}
\acmBooktitle{Extended Abstracts of the CHI Conference on Human Factors in Computing Systems (CHI EA '25), April 26-May 1, 2025, Yokohama, Japan}\acmDOI{10.1145/3706599.3719759}
\acmISBN{979-8-4007-1395-8/2025/04}

\usepackage{multirow}
\usepackage{framed,enumitem} 
\usepackage{graphicx}
\usepackage{xspace}
\usepackage{makecell}
\usepackage{colortbl}
\usepackage{tabularx}
\usepackage{tikz}

\definecolor{pastelblue}{HTML}{D2F0FE}
\definecolor{pastelgreen}{HTML}{DFF8E4}
\definecolor{pastelpink}{HTML}{FFE4E9}
\definecolor{pastelyellow}{HTML}{FFFAE5}

\definecolor{stage1color}{HTML}{F17321}
\definecolor{stage2color}{HTML}{80B728}
\definecolor{stage3color}{HTML}{179D6C}

\newcommand{\qualitybox}[1]{%
  \tikz[baseline=(X.base)] \node[fill=pastelpink, rounded corners=3pt, inner sep=3pt] (X) {\textbf{#1}};}
\newcommand{\quantitybox}[1]{%
  \tikz[baseline=(X.base)] \node[fill=pastelyellow, rounded corners=3pt, inner sep=3pt] (X) {\textbf{#1}};}
\newcommand{\relationbox}[1]{%
  \tikz[baseline=(X.base)] \node[fill=pastelgreen, rounded corners=3pt, inner sep=3pt] (X) {\textbf{#1}};}
\newcommand{\mannerbox}[1]{%
  \tikz[baseline=(X.base)] \node[fill=pastelblue, rounded corners=3pt, inner sep=3pt] (X) {\textbf{#1}};}

\begin{document}

\title[Applying the Gricean Maxims to a Human-LLM Interaction Cycle]{Applying the Gricean Maxims to a Human-LLM Interaction Cycle: Design Insights from a Participatory Approach}

\author{Yoonsu Kim}
\email{yoonsu16@kaist.ac.kr}
\affiliation{%
  \institution{School of Computing, KAIST}
  \city{Daejeon}
  \country{Republic of Korea}
}
\author{Brandon Chin}
\email{brandoncjw@hkn.eecs.berkeley.edu}
\affiliation{%
  \institution{University of California, Berkeley}
  \city{Berkeley}
  \country{United States of America}
}

\author{Kihoon Son}
\email{kihoon.son@kaist.ac.kr}
\affiliation{%
  \institution{School of Computing, KAIST}
  \city{Daejeon}
  \country{Republic of Korea}
}
\author{Seoyoung Kim}
\email{youthskim@kaist.ac.kr}
\affiliation{%
  \institution{School of Computing, KAIST}
  \city{Daejeon}
  \country{Republic of Korea}
}

\author{Juho Kim}
\email{juhokim@kaist.ac.kr}
\affiliation{%
  \institution{School of Computing, KAIST}
  \city{Daejeon}
  \country{Republic of Korea}
}

\renewcommand{\shortauthors}{Yoonsu Kim et al.}




\begin{teaserfigure}
  \includegraphics[width=\textwidth]{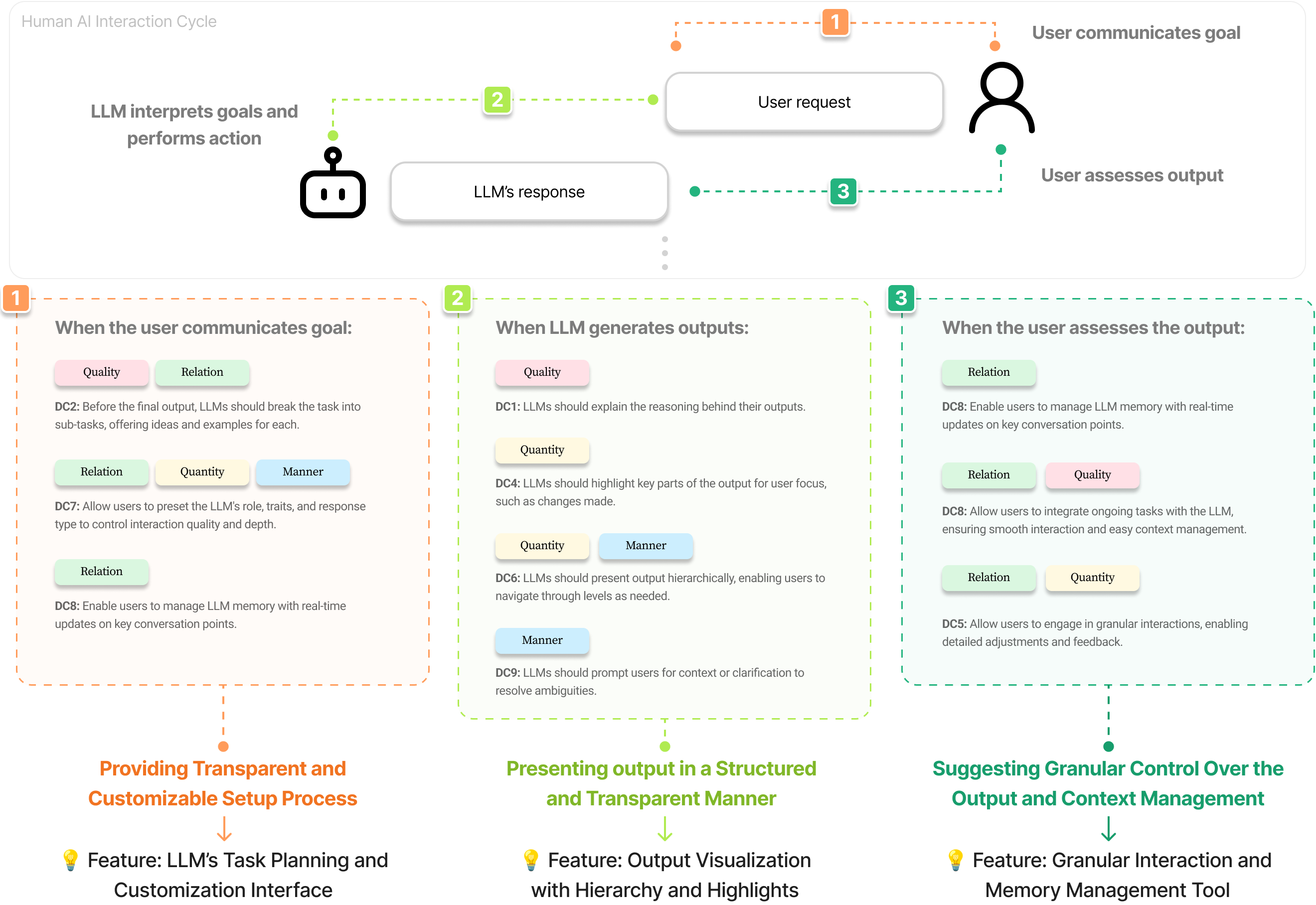}
  \Description{This figure presents nine design considerations (DCs) grounded in the Gricean Maxims, categorized into three stages of the Human-AI interaction cycle: (1) User communicates a goal, (2) LLM interprets the goal and performs actions, and (3) User assesses the output. Each stage includes specific DCs tailored to the goals of that phase, such as task planning and customization during goal communication, hierarchical output visualization with highlights during LLM output generation, and granular interaction tools and real-time memory management during output assessment.}
  \caption{We grouped a total of nine design considerations (DCs) grounded in the Gricean Maxims according to the three stages of the Human-AI interaction cycle: (1) User communicates a goal, (2) LLM interprets the goal and performs actions, and (3) User assesses the output. Based on the common goals of each stage, we propose corresponding design features.}
  \label{fig:design-goal}
\end{teaserfigure}

\begin{abstract}
While large language models (LLMs) are increasingly used to assist users in various tasks through natural language interactions, these interactions often fall short due to LLMs' limited ability to infer contextual nuances and user intentions, unlike humans. To address this challenge, we draw inspiration from the Gricean Maxims---human communication theory that suggests principles of effective communication---and aim to derive design insights for enhancing human-AI interactions (HAI). Through participatory design workshops with communication experts, designers, and end-users, we identified ways to apply these maxims across the stages of the HAI cycle. Our findings include reinterpreted maxims tailored to human-LLM contexts and nine actionable design considerations categorized by interaction stage. These insights provide a concrete framework for designing more cooperative and user-centered LLM-based systems, bridging theoretical foundations in communication with practical applications in HAI.
\end{abstract}
\maketitle
\section{Introduction}

With the recent advancements in large language models (LLMs), people prevalently engage in linguistic communication with LLMs, having conversations to accomplish various tasks~\cite{ChatGPT:online, Gemini:online}. 
Linguistic communication, by nature, is not a mere succession of disconnected remarks; rather, it is a cooperative process in which participants work toward a shared purpose or mutually accepted direction~\cite{grice1975logic, mey2001pragmatics, kasirzadeh2023conversation, clark1983common, clark1981definite}. Similarly, the linguistic communication between users and LLMs also requires a collaborative effort with a mutual direction: successfully accomplishing the user's task.
However, the cooperative nature of such interactions often falls short. One reason is that, unlike humans, LLMs lack inferential abilities that allow them to understand contextual nuances or interpret others' intentions~\cite{kasirzadeh2023conversation}. This often leads to diminished task performance, reduced user satisfaction, and sometimes users abandoning the interaction~\cite{qian2024tell, kim2023understanding, bodonhelyi2024user}.

Philosopher H.P. Grice formalized the cooperative nature of communication through \textbf{\textit{Gricean Maxims}}, which assert that effective communication relies on interlocutors adhering to certain maxims: \quantitybox{Quantity} (be informative), \qualitybox{Quality} (be truthful), \relationbox{Relation} (be relevant), and \mannerbox{Manner} (be clear)~\cite{grice1975logic}. These maxims provide a foundational framework not only for understanding how people communicate cooperatively but also for understanding communication between humans and machines in NLP and human-AI interaction~\cite{krause-vossen-2024-gricean-maxims, Panfili_Duman_Nave_Ridgeway_Eversole_Sarikaya_2021}. 
For example, NLP researchers employed the Gricean Maxims as a framework for evaluating the capabilities of models like LLMs based on their adherence to these maxims~\cite{khayrallah-sedoc-2021-measuring, miehling2024language}. In human-AI interaction research, studies have applied the Gricean Maxims to real-world scenarios, such as commercial chatbots, to test their effectiveness in terms of each maxim~\cite{Panfili_Duman_Nave_Ridgeway_Eversole_Sarikaya_2021, xiao2020tell, setlur2022analyticalchatbot}.

However, there has been little exploration of how the Gricean Maxims can be practically applied across the overall human-AI interaction cycle by combining both perspectives of domain experts (e.g., communication and HAI design) and end-users.
Exploring this can provide valuable guidance on how HAI systems should be designed to support people better, not only reflecting the theory but also addressing the real-world complexities of human-LLM interaction, ensuring a balance between theoretical foundations, practical design considerations, and real user needs.
Therefore, we conducted participatory design workshops involving participants from three fields---communication experts, interface/interaction designers, and experienced end-users---to gather design ideas for applying the Gricean Maxims to the human-AI interaction cycle. The reason for choosing participatory design workshops, a method widely used in HCI, is their effectiveness in integrating diverse perspectives to derive actionable design insights~\cite{zytko2022participatory}.

Through a series of four workshops with 10 participants, we gathered qualitative data, including workshop transcriptions and collaboratively developed design ideas. We qualitatively analyzed the data and derived a total of 9 design considerations to improve human-LLM interactions grounded in the Gricean Maxims (\autoref{tab:design-consideration}). We also identified that participants redefined the Gricean Maxims within the context of human-LLM interactions (Section ~\ref{ref:redefinition}). Furthermore, we grouped these design considerations into the stages of the human-AI (LLM) interaction cycle: (1) users communicate their goals, (2) the AI interprets goals and performs actions (LLM generates the output), and (3) users assess the output~\cite{terry2023ai} to provide more actionable and practical design implications. This approach led us to identify the primary design considerations for each stage and propose corresponding design features tailored to these considerations (\autoref{fig:design-goal}, Section~\ref{ref:design-implication}). Our findings provide concrete ideas for advancing human-LLM interaction by leveraging insights from human communication theory, paving the way for more cooperative and user-centric LLM-based systems.
\section{Background}

\subsection{Cooperative Principle and Gricean Maxims}
Gricean Maxims, formulated by H.P. Grice as part of his Cooperative Principle, describe the traits of successful cooperative communication: \quantitybox{Quantity}, \qualitybox{Quality}, \relationbox{Relation}, and \mannerbox{Manner} ~\cite{grice1975logic}. In terms of \quantitybox{Quantity}, participants should provide the necessary amount of information, without offering more than needed. The \qualitybox{Quality} maxim emphasizes truthfulness, advising against sharing false information or claims lacking evidence. \relationbox{Relation} maxim focuses on relevance, urging speakers to provide information pertinent to the exchange and omit irrelevant details. The \mannerbox{Manner} maxim stresses clarity, advocating for avoiding obscurity and ambiguity while promoting brevity and orderliness.
Gricean Maxims describe effective human communication strategies, providing a theoretical framework that has influenced the development of NLP and HCI technologies~\cite{krause-vossen-2024-gricean-maxims, Panfili_Duman_Nave_Ridgeway_Eversole_Sarikaya_2021}. In particular, with the advancement of LLM-based systems designed for natural language conversations, incorporating the principles into their design could lead to meaningful improvement. Motivated by this, we aim to explore how the Gricean Maxims can inform the design of human-LLM interactions, specifically at each stage of the human-LLM interaction cycle.

\subsection{Gricean Maxim in Human-AI Interaction}
As foundational principles in pragmatics, Gricean Maxims have been widely applied and tested in the context of human-AI interactions, offering insights into how effective communication can be achieved. 
Research has shown that adherence to these maxims improves engagement, response quality, and user satisfaction in chatbot interactions~\cite{xiao2020tell, setlur2022analyticalchatbot}, while violations---particularly of \relationbox{Relation} maxim---often result in user frustration ~\cite{Panfili_Duman_Nave_Ridgeway_Eversole_Sarikaya_2021, nam2023language}. These findings highlight the importance of incorporating the Gricean Maxims to guide the design of conversational agents.
However, recent studies highlight that directly applying the Gricean Maxims to human-AI interaction is not straightforward due to AI's difficulty in adapting nuanced contextual variation~\cite{kasirzadeh2023conversation}. Furthermore, researchers have augmented the original maxims with additional principles such as \textit{\textbf{Benevolence}} (moral responsibility) and \textit{\textbf{Transparency}} (recognition of knowledge boundaries and constraints) for addressing unique challenges in human-AI interaction~\cite{miehling2024language}. 
While these studies highlight the potential of Gricean Maxims in shaping effective conversational agents, their practical application within the human-AI interaction cycle, integrating perspectives from both domain experts (e.g., communication and HAI design) and end-users, remains underexplored.
To address this, we conducted participatory design workshops to gather theoretical, practical, and user-centered design ideas.
\section{Participatory Design Workshop}
The workshop's objective was to gather design ideas and considerations for applying each maxim to improve human-LLM interactions.
By involving a diverse group of experts, including a communication expert, an interface/interaction designer, and an experienced LLM user as one team, we seek to ensure that the ideas are theoretically informed, practically feasible, and user-centric. The following sections outline the recruitment process, participant details, workshop protocols, and the analysis process of the workshop data.

\subsection{Recruitment}
In our workshop, we invited two types of experts (communication experts and interface/interaction designers) along with experienced LLM users. 
The rationale for recruiting participants from these three distinct fields, their specific roles and responsibilities in the workshop, and the criteria for their selection are detailed below:

\textbf{Communication Experts}
The inclusion of communication experts is essential because they provide in-depth knowledge of the communication theory (i.e., Gricean Maxims) and its application in various communication contexts. We recruited Ph.D. holders or Ph.D. students majoring in communication for this role. In the workshop, they were asked to provide guidance to ensure that the devised design ideas aligned with the Gricean Maxims. 

\textbf{Interface/Interaction Designers}
We recruited interface and interaction designers who have experience in designing interfaces or interactions that involve AI, particularly LLMs. To demonstrate their expertise, they were required to submit design outcomes they had worked on that involved LLMs or AI. They play a critical role in translating theoretical concepts into practical design solutions. Their expertise in designing interfaces and interactions for AI systems enables them to propose feasible and innovative design ideas. By presenting examples of existing AI-based systems, they can help bridge the gap between theory and practice. 

\textbf{Experienced LLM Users}
To incorporate end-user perspectives into the design process, we invited experienced LLM users who frequently used LLMs (at least four times a week) to our workshop. Before the workshop, they were asked to submit at least two recent conversations with LLMs via shared links~\footnote{https://help.openai.com/en/articles/7925741-chatgpt-shared-links-faq, https://support.google.com/gemini/answer/13743730}. We also asked them to provide information on their goal of LLM usage, specific tasks they were trying to accomplish, and any dissatisfaction they experienced during the interaction. We selected conversations covering various tasks, which included academic writing, content refinement, drafting, and curriculum design.
These conversations were used in our workshop to facilitate the brainstorming of design ideas in real-world scenarios. 
Their role was crucial in providing end-user perspectives, sharing specific conversation cases, and helping the experts understand the practical implications of the proposed design ideas.

\subsection{Participants}
Each workshop included three participants, excluding the facilitator and assistants: one communication expert, one interface/interaction designer, and one experienced LLM user. We conducted a total of four workshops, recruiting 10 participants in total: two communication experts, four interaction/interface designers, and four experienced LLM users.
The two communication experts participated in two workshops each. We asked them to join multiple sessions because their role was to provide guidance to ensure that the devised design ideas were aligned with the Gricean Maxims, which minimized the risk of introducing bias during the ideation process. Before the workshops, we provided the communication experts with tutorial materials on the Gricean Maxims, which they were required to review beforehand.
The average age of participants was 28 years (SD = 2.94), with 3 females and 7 males. During recruitment, we surveyed participants' knowledge of LLMs and their frequency of use through self-reporting. All participants reported using LLMs at least four times per week, indicating that they were all frequent users of LLMs. Details about participant information can be found in ~\autoref{tab:workshop-participants}.
For compensation, communication experts and interface/interaction designers, as professional participants, were paid 80,000 KRW (Approx. 60 USD) per hour. Experienced LLM users, although not professionals, received 60,000 KRW (Approx. 45 USD) per hour, reflecting the additional requirement of submitting a prior conversation for workshop use.

\subsection{Workshop Protocol}
The workshop was conducted online via Zoom~\footnote{https://zoom.us/} and utilized FigJam~\footnote{https://www.figma.com/figjam/} for collaborative activities. The total duration of the workshop was two hours, structured into four main sessions as follows. An example of the FigJam board used in the workshop can be found in ~\autoref{fig:figjam-board}.

\textbf{Session 1. Introduction to the research and tools (10 mins)}
The workshop began with an introduction to our research objectives and a brief tutorial on the tool (FigJam) that will be used throughout the workshop. Participants also briefly introduced themselves to foster a collaborative environment. 

\textbf{Session 2. Introduction to communication theory, Gricean maxims (20 mins)}
Following the introduction, one of the authors explained the Gricean maxims (\qualitybox{Quality}, \quantitybox{Quantity}, \relationbox{Relation}, \mannerbox{Manner}). Examples of good and bad adherence to each maxim were provided to clarify their practical application.

\textbf{Session 3. Analysis of LLM conversations and brainstorming of design ideas (60 mins)}
The primary goal of this session was to analyze LLM conversations submitted in advance by LLM users and brainstorm design ideas to enhance adherence to the Gricean Maxims.
In this session, the LLM user first introduced the conversation, explained its purpose, and shared their retrospective experiences for each turn. Then, the participants engaged in a turn-by-turn analysis of the conversation and brainstormed design ideas for each maxim. This session was divided into three sub-sessions. \textbf{
(1) Free Brainstorming}: Participants reviewed the first two turns of conversation, identified areas for improvement based on the four maxims, and brainstormed and documented design ideas on FigJam. \textbf{(2) Getting References}: The authors provided and explained some examples of design ideas from existing LLM-based systems~\cite{NotionAI53:online, AIWritin58:online, Quillbot:online, Replit:online, Gemini:online} and literature~\cite{aiscribe2024, explorellm2024, wu2022promptchainerchaininglargelanguage, sensecape2023, graphologue2023, mirror2023, chainforge2024} to the participants. \textbf{(3) Guided Brainstorming with References}: Participants continued the turn-by-turn review, using the references as inspiration for brainstorming additional design ideas.

\textbf{Session 4. Reviewing, Elaborating, and Integrating Design ideas (30 mins)}
In this final session, the focus shifted to reviewing and discussing the proposed design ideas, elaborating on their details, and discussing how they can be applied to a broader range of conversational contexts beyond the specific scenarios discussed in the workshop. Participants were encouraged to draw on their own past experiences with LLMs to further develop these ideas, ensuring they were adaptable to various situations. This process involved critically evaluating and elaborating on the proposed ideas, with the option to create UI sketches to visually represent and refine the designs.

\subsection{Workshop Analysis}
The workshop sessions lasted an average of 119.75 minutes (SD: 2.63). On average, analyzed conversations had 23.5 turns (SD: 9.3). All sessions were conducted over Zoom and recorded with participants' prior consent, following approval from our institution's IRB.
All design ideas generated during the workshops, along with corresponding session recording transcripts, were documented. 
The initial step of our analysis involved compiling a list of all design ideas and their targeted maxim(s) from the four workshops. Using the recording transcripts, we supplemented this list with participants’ explanations of how each design idea could support the corresponding maxim(s) within the context of human-AI interaction. This resulted in a total of 102 design ideas.

Two authors conducted a qualitative analysis of the 102 design ideas using open and axial coding~\cite{gibbs2018analyzing}.
In the initial coding phase, we independently coded the ideas, focusing on how they supported adherence to their targeted maxim(s).
Through this inductive process, we created a set of codes that captured the distinct characteristics of each design idea to support the corresponding maxim(s).
Building on this, we grouped the codes into higher-level categories, identifying common design considerations shared across multiple ideas. Through iterative discussions in this process, we derived nine key design considerations for enhancing human-LLM interaction grounded in the Gricean Maxims (\autoref{tab:design-consideration}).
Additionally, the analysis revealed that participants had reinterpreted the Gricean Maxims, originally described for human-human communication, to align with the specific context of human-LLM interactions. By examining how multiple design ideas targeting the same maxim were grouped, we uncovered their reinterpretations (Section~\ref{ref:redefinition}).
Furthermore, to make these insights more tangible, we mapped the nine design considerations to the three distinct stages of the human-AI interaction cycle, offering a structured framework for their practical application (Section~\ref{ref:design-implication}).
\section{Workshop Findings}
This section presents our workshop findings, including the reinterpretation of the Gricean Maxims, nine design considerations, and their grouping within the human-AI interaction cycle.

\subsection{Reinterpretation of Gricean Maxims in the context of human-LLM interaction}~\label{ref:redefinition}

Through our analysis, we identified how each maxim could be reinterpreted to better align with human-LLM interaction, reflecting the unique dynamics of human-LLM communication compared to human-human communications.

\quantitybox{Maxim of Quantity}, originally defined as making contributions as informative as necessary, was reinterpreted to include optimizing the user's cognitive load. 
This involves presenting information in a hierarchical manner, allowing users to start with a summary and explore more details as needed, avoiding both excessive and inadequate information. 

\qualitybox{Maxim of Quality}, originally stressed providing truthful and evidence-based contributions, was expanded to include fostering user trust in the LLMs. Participants highlighted that, unlike human partners, the various limitations of current LLMs, such as hallucinations and inconsistencies, undermine their perceived reliability. Therefore, a trust-building aspect should be included in human-LLM interactions for users to perceive LLMs as reliable partners. This involves not only providing accurate information but also ensuring users can easily recognize its accuracy and reasoning process.

\relationbox{Maxim of Relation}, originally about ensuring the relevance of information in the current exchange, was refined to emphasize how LLMs manage the user's broader context throughout the conversation. This approach goes beyond merely addressing the immediate query and involves following the overall conversation's flow and adapting dynamically to the user's evolving goals and intentions. 

Lastly, the \mannerbox{Maxim of Manner}, which traditionally advocates for clarity, brevity, and orderliness, was reinterpreted to emphasize the importance of the LLM adapting its output format to meet the user’s specific needs. Given the diversity of user tasks using LLMs and the variety of user preferences, LLMs should tailor their output to match the user's specific needs---whether through bullet points, summaries, detailed explanations, or specialized formats like code, etc. This flexibility ensures that interactions remain clear, effective, and aligned with user expectations.

\begin{table*}[ht]
\renewcommand{\arraystretch}{1.2}
\small
\centering
\begin{tabularx}{\textwidth}{>{\centering\arraybackslash}m{0.7cm}|
                                    >{\centering\arraybackslash}m{1.5cm}|
                                    >{\centering\arraybackslash}X|
                                    >{\centering\arraybackslash}m{4.5cm}}
\toprule
\textbf{\#} & \textbf{Targeted Maxim} & \textbf{Design Considerations} & \textbf{Stage} \\
\midrule
DC1 & \qualitybox{Quality} & \textbf{LLMs should provide the reasoning and explanations behind the LLM's generated outputs.} & \textbf{\textcolor{stage2color}{(2) The LLM generates the output}} \\ \hline
DC2 & \makecell{\qualitybox{Quality}, \\ \relationbox{Relation}} & \textbf{Before the final output, LLMs should present how it will perform the user’s task by decomposing it into sub-tasks, providing relevant ideas and examples for each.} & \textbf{\textcolor{stage1color}{(1) Users communicate their goals}} \\ \hline
DC3 & \makecell{\qualitybox{Quality}, \\ \relationbox{Relation}} & \textbf{Enable users to seamlessly integrate their ongoing tasks with the LLM, allowing for uninterrupted interaction while easily managing the necessary context.} & \textbf{\textcolor{stage3color}{(3) Users assess the output}} \\ \hline
DC4 & \quantitybox{Quantity} & \textbf{LLMs should present its output by highlighting specific parts that the user should focus on (e.g., which parts were changed).} & \textbf{\textcolor{stage2color}{(2) The LLM generates the output}} \\ \hline
DC5 & \makecell{\quantitybox{Quantity}, \\ \relationbox{Relation}} & \textbf{Enable users to engage in more granular interactions with the LLM, allowing them to modify, adjust, and provide feedback at a detailed level.} & \textbf{\textcolor{stage3color}{(3) Users assess the output}} \\ \hline
DC6 & \makecell{\quantitybox{Quantity}, \\ \mannerbox{Manner}} & \textbf{LLMs should organize and present their output in a hierarchical manner, allowing users to navigate through each level of the structure as needed.} & \textbf{\textcolor{stage2color}{(2) The LLM generates the output}} \\ \hline
DC7 & \makecell{\quantitybox{Quantity}, \\ \relationbox{Relation}, \\ \mannerbox{Manner}} & \textbf{Enable users to preset the LLMs role, characteristics, and the type of responses they seek, enabling them to control both quantitative and qualitative aspects of the interaction.} & \textbf{\textcolor{stage1color}{(1) Users communicate their goals}} \\ \hline
DC8 & \relationbox{Relation} & \textbf{Enable users to easily monitor and manage LLM's memory (what the LLM remembers, forgets, or updates) by providing dynamic updates on key points of the conversation or LLM's memory throughout the conversation.} & \makecell{\textbf{\textcolor{stage1color}{(1) Users communicate their goals}} \\ \textbf{\textcolor{stage3color}{(3) Users assess the output}}} \\ \hline
DC9 & \mannerbox{Manner} & \textbf{LLMs should guide users to provide additional context or clarification to address ambiguity or difficulties when generating responses.} & \textbf{\textcolor{stage2color}{(2) The LLM generates the output}} \\ 
\bottomrule
\end{tabularx}
\caption{A total of 9 design considerations for enhancing human-LLM interaction grounded in Gricean Maxims. Each consideration targets specific maxims and is mapped to the three stages of the human-AI interaction cycle: \textcolor{stage1color}{(1) Users communicate their goals}, \textcolor{stage2color}{(2) The LLM generates the output}, \textcolor{stage3color}{(3) Users assess the output}.}
\label{tab:design-consideration}
\end{table*}

\subsection{Design considerations for enhancing human-LLM interaction grounded in Gricean Maxims}
Building on these redefined maxims, we identified nine design considerations for enhancing human-LLM interaction through qualitative analysis of our workshop data (Table ~\ref{tab:design-consideration}). 
Each consideration was mapped to its corresponding maxims and categorized into the stages of the human-AI interaction cycle. In this section, we elaborate on how each consideration aligns with its targeted maxim(s) by helping adhere to them and improving human-LLM interaction.

\textbf{DC1} supports the \qualitybox{Quality} by allowing users to assess the reliability of the information and better understand the logic behind the responses.
\textbf{DC2} ensures that the output remains accurate and relevant to the user’s goals and context, adhering to the \relationbox{Relation}, while also targeting the \qualitybox{Quality} by allowing users to review and verify the output before completion, thereby enhancing trust.
\textbf{DC3} helps provide accurate and contextually appropriate responses, supporting the \qualitybox{Quality} and the \relationbox{Relation}.
\textbf{DC4} helps reduce cognitive load and enables users to effectively identify the most important information, supporting the \quantitybox{Quantity}.
\textbf{DC5} supports the \quantitybox{Quantity} by giving users precise control over the level of detail, while maintaining relevance in the interaction, which adheres to the \relationbox{Relation}.
\textbf{DC6} ensures clarity and structure in the presentation of information, supporting both the \quantitybox{Quantity} and \mannerbox{Manner}.
\textbf{DC7} supports the \mannerbox{Manner} by allowing users to clearly define the desired response format, the \relationbox{Relation} by ensuring relevance to user preferences and context, and the \quantitybox{Quantity} by helping users control the amount of information they receive.
\textbf{DC8} helps maintain coherence and alignment with the user’s goals, adhering to the \relationbox{Relation}.
\textbf{DC9} supports the \mannerbox{Manner} by resolving ambiguities and ensuring clear and effective communication.

\subsection{Mapping design considerations into a human-AI interaction cycle}~\label{ref:design-implication}
To gain a more concrete and practical understanding of how the nine design considerations can be applied and realized within the human-LLM interaction process, we mapped them onto the three stages of the human-AI (LLM) interaction cycle: (1) users communicate their goals, (2) the AI interprets goals and performs actions (LLM generates the output), and (3) users assess the output~\cite{terry2023ai}. By situating these considerations within this cycle, we aimed to clarify their applicability at each stage of interaction.
Additionally, by mapping the design considerations to each stage of the cycle, we identified common objectives at each stage. Based on these common goals, we proposed actionable design features for each phase (Figure~\ref{fig:design-goal}).

\noindent
\textcolor{stage1color}{\textbf{(1) When users communicate their goals}} In this stage, a common objective is to \textbf{provide a transparent and customizable setup process}. This includes allowing users to see how the LLM will perform their tasks, set up its role, and manage its memory, ensuring that the LLM’s understanding and execution process are clearly communicated. 
A specific design feature for this stage is an interface in which the LLM proposes a detailed plan on how it will approach the user's task by decomposing it into sub-tasks and suggesting its role, characteristics, response type, etc., which the user can approve or modify to ensure alignment with their expectations. 

\noindent
\textcolor{stage2color}{\textbf{(2) When the LLM generates the output}}
In this stage, the common goal is to \textbf{present the outputs in a structured and transparent way}. This involves offering reasoning and explanations for the generated answers, highlighting focus areas, and enabling users to navigate through varying levels of detail. 
A key design feature for this stage is presenting LLM-generated outputs in a hierarchical, expandable/collapsible format. Key sections can be highlighted, and users can have access to on-demand explanations of the LLM's reasoning, along with prompts for further context as needed. 

\noindent
\textcolor{stage3color}{\textbf{(3) When users assess the output}} The common goal in this stage is to \textbf{provide users with granular control over the output and context management} capabilities, allowing users to engage with the output at a detailed level, maintain alignment with their ongoing tasks, and ensure the coherence of the conversation by managing the LLM’s memory effectively.
A proposed design feature for this stage is a tool that allows users to interact with, modify, and provide feedback on specific sections of the LLM’s output or LLM’s memory status. This tool should seamlessly integrate with users' ongoing tasks to manage context and updates without interruption.
\section{Limitation and Future Work}
One limitation of this study is that the proposed design considerations have not yet undergone empirical evaluation. While these considerations are grounded in theoretical insights and encompass a variety of participant perspectives, their feasibility and effectiveness remain invalidated. 
Future work should focus on systematically evaluating these design features through user studies, controlled experiments, or longitudinal assessments to measure their impact on usability, communication efficiency, and user satisfaction. Iterative prototyping and task-based user studies will help assess how effectively these features enhance user satisfaction and task performance in real-world scenarios.

Additionally, we acknowledge that some of the proposed features may already be partially realized as commercial products or research prototypes. Therefore, future research could include a systematic analysis of existing work in HCI and NLP to evaluate the extent to which the design considerations proposed in this study are addressed in current approaches to human-LLM interaction. By identifying gaps in coverage or areas of overlap, this analysis could help refine the applicability of these considerations and highlight opportunities for further innovation.
By addressing these limitations and extending the findings, future research can contribute to the development of conversational agents that are not only more cooperative and intuitive but also more broadly applicable across diverse contexts and user needs.
\begin{acks}
This work was supported by the Institute of Information \& Communications Technology Planning \& Evaluation (IITP) grant funded by the Korean government (MSIT) (No.2021-0-01347, Video Interaction Technologies Using Object-Oriented Video Modeling).
This work was supported by Institute of Information \& communications Technology Planning \& Evaluation (IITP) grant funded by the Korea government(MSIT) (No. RS-2024-00443251, Accurate and Safe Multimodal, Multilingual Personalized AI Tutors).
\end{acks} 

\bibliographystyle{ACM-Reference-Format}
\bibliography{references}
\section{appendix}
\subsection{Workshop participant details}
Details about participant information can be found in ~\autoref{tab:workshop-participants}.
\begin{table*}[h]
\resizebox{\textwidth}{!}{%
\small
\begin{tabular}{c|c|c|c|c|c}
\toprule
\textbf{Role} & \textbf{\makecell[c]{Workshop \\ Session}} & \textbf{Age} & \textbf{Gender} & \textbf{Expertise} & \textbf{Knowledge of LLM} \\ 
\midrule
\multirow{2}{*}{\textbf{\makecell[c]{Communication \\ Expert}}} & W1, W2 & 35 & Female & Ph.D. in Communication & Knowledgeable \\ \cline{2-6} 
 & W3, W4 & 27 & Female & Ph.D. student in Communication & Neutral \\ \hline
\multirow{4}{*}{\textbf{\makecell[c]{Interface/Interaction \\ Designer}}} & W1 & 25 & Male & M.S. student in Industrial Design & Neutral \\ \cline{2-6} 
 & W2 & 28 & Male & Ph.D. in Industrial Design & Knowledgeable \\ \cline{2-6} 
 & W3 & 28 & Male & Ph.D. student in Industrial Design & Knowledgeable \\ \cline{2-6} 
 & W4 & 29 & Female & Ph.D. student in Industrial Design & Neutral \\ \hline
\multirow{4}{*}{\textbf{LLM user}} & W1 & 27 & Male & \multicolumn{1}{c|}{-} & Very knowledgeable \\ \cline{2-6} 
 & W2 & 29 & Male & \multicolumn{1}{c|}{-} & Knowledgeable \\ \cline{2-6} 
 & W3 & 24 & Male & \multicolumn{1}{c|}{-} & Knowledgeable \\ \cline{2-6} 
 & W4 & 28 & Male & \multicolumn{1}{c|}{-} & Low knowledgeable \\ 
 \bottomrule
\end{tabular}%
}
\caption{Workshop participants' demographics, expertise, and their knowledge level of LLM.}
\label{tab:workshop-participants}
\end{table*}

\subsection{An example of a FigJam board used in our workshop}
\begin{figure*}
  \begin{minipage}{\textwidth}
  \resizebox{\textwidth}{!}{\includegraphics{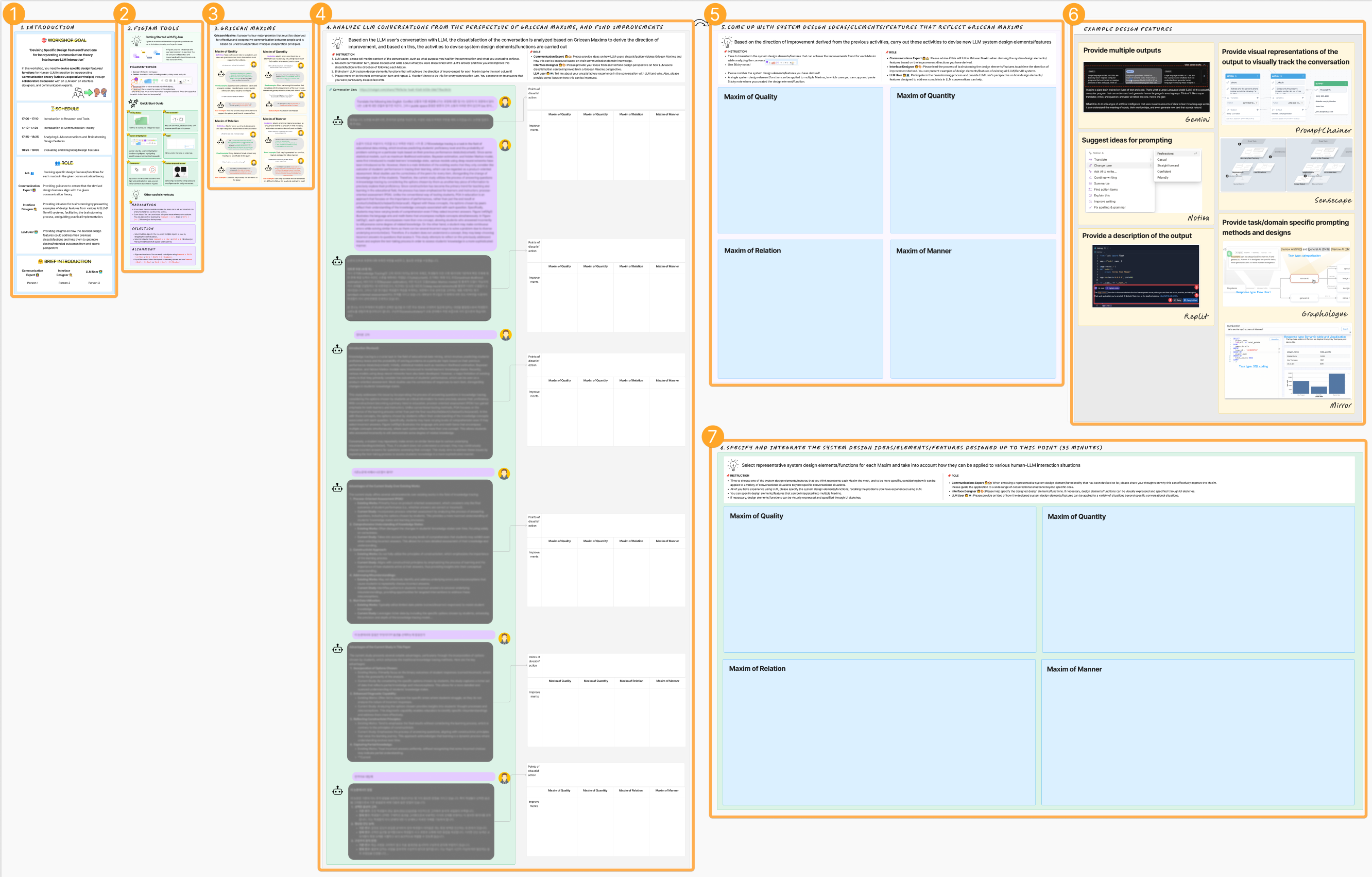}}
  \caption{An example of a FigJam Board that we used in our participatory design workshop. It includes (1) \textbf{Introduction} of the workshop and its objectives; (2) \textbf{Tutorial} on FigJam, the collaborative tool used for the workshop; (3) \textbf{Explanation of Gricean Maxims}; (4) \textbf{LLM conversation analysis}, which is pre-submitted from the LLM user, identifying areas for improvement based on Gricean Maxims (with conversation content blurred for privacy protection); (5) \textbf{Brainstorming design Ideas}, (6) \textbf{Example design features}, which are provided after deriving a few design ideas from analyzing two turns of conversation; (7) \textbf{Refining and combining design ideas} into actionable solutions.}
  \label{fig:figjam-board}
  \end{minipage}
\end{figure*}

An example of the FigJam board we used in our workshop can be found in ~\autoref{fig:figjam-board}.

\end{document}